\documentclass[conference]{IEEEtran}
\usepackage[bookmarks=false]{hyperref}
\IEEEoverridecommandlockouts
\usepackage{cite}
\usepackage{csquotes}
\usepackage{graphicx}
\usepackage{amsmath,amssymb,amsfonts}
\usepackage{algorithmic}
\usepackage{textcomp}
\usepackage{xcolor}
\usepackage{amsmath}
\usepackage{multirow}
\usepackage{authblk}
\usepackage{siunitx}
\usepackage{subcaption}
\usepackage{comment}
\makeatletter
\def\footnoterule{\kern-3\p@
  \hrule \@width 2in \kern 2.6\p@} % the \hrule is .4pt high
\makeatother
\usepackage{lipsum}
\makeatletter
\def\hyper@refstepcounter#1{%
  \edef\This@name{#1}%
  \ifx\This@name\name@of@eq
    \@ifundefined{theHequation}{%
      \make@stripped@name{\theequation}%
      \let\theHequation\newname
    }{}%
  \fi
  \HyCnt@ProvideTheHCounter{#1}%
  \hyper@makecurrent{#1}%
  \ifmeasuring@
  \else
    \Hy@raisedlink{%
      \hyper@anchorstart{\@currentHref}\hyper@anchorend
    }%
  \fi
  \ifmmode\mathopen{}\fi           %% second version: THIS LINE ADDED IN PATCH
}
\makeatother
\usepackage[style=base,font=small,labelfont=bf]{caption}
\newcommand{\RNum}[1]{\uppercase\expandafter{\romannumeral #1\relax}}

\def\BibTeX{{\rm B\kern-.05em{\sc i\kern-.025em b}\kern-.08em
    T\kern-.1667em\lower.7ex\hbox{E}\kern-.125emX}}
\begin{document}

\title{DeepAtom: A Framework for Protein-Ligand Binding Affinity Prediction\\
}
\author[1,\dag]{Yanjun Li\thanks{\dag These authors contributed equally.}}
\author[2,3,\dag]{Mohammad A. Rezaei}
\author[2,*]{Chenglong Li}
\author[1,*]{Xiaolin Li\thanks{* To whom correspondence should be addressed.}}
\author[1]{Dapeng Wu}
\affil[1]{\small NSF Center for Big Learning, University of Florida}
\affil[2]{\small Department of Medicinal Chemistry, Center for Natural Products, Drug Discovery and Development (CNPD3), University of Florida}
\affil[3]{\small Department of Chemistry, University of Florida.}
\affil[ ]{\small Gainesville, FL, USA}
\affil[ ]{\small yanjun.li@ufl.edu, mohammadarezaei@chem.ufl.edu, lic@cop.ufl.edu, andyli@ece.ufl.edu, dpwu@ufl.edu}
\IEEEoverridecommandlockouts
% \IEEEpubid{\makebox[\columnwidth]{978-1-7281-1867-3/19/\$31.00 ©2019 IEEE \hfill} \hspace{\columnsep}\makebox[\columnwidth]{ }}
\maketitle
% \IEEEpubidadjcol
% https://site.ieee.org/compel2018/ieee-copyright-notice/

\begin{abstract}
The cornerstone of computational drug design is the calculation of binding affinity between two biological counterparts, especially a chemical compound, i.e., a ligand, and a protein. 
Predicting the strength of protein-ligand binding with reasonable accuracy is critical for drug discovery. 
In this paper, we propose a data-driven framework named DeepAtom to accurately predict the protein-ligand binding affinity. 
With 3D Convolutional Neural Network (3D-CNN) architecture, DeepAtom could automatically extract binding related atomic interaction patterns from the voxelized complex structure.
Compared with the other CNN based approaches, our light-weight model design effectively improves the model representational capacity, even with the limited available training data.
With validation experiments on the PDBbind v.\num{2016} benchmark and the independent Astex Diverse Set, we demonstrate that the less feature engineering dependent DeepAtom approach consistently outperforms the other state-of-the-art scoring methods.
We also compile and propose a new benchmark dataset to further improve the model performances. 
With the new dataset as training input, DeepAtom achieves Pearson\textquotesingle s R=\num{0.83} and RMSE=\num{1.23} $pK$ units on the PDBbind v.\num{2016} core set. 
The promising results demonstrate that DeepAtom models can be potentially adopted in computational drug development protocols such as molecular docking and virtual screening.

\end{abstract}


\section{Introduction}
Binding of a chemical molecule to a protein may start a biological process. 
This includes the activation or inhibition of an enzyme's activity, and a drug molecule affecting its target protein. 
The binding is quantified by how strong the chemical compound, a.k.a. a ligand, binds to its counterpart protein; this quantity is called \textit{binding affinity}. 
Simulation of biological processes and computational drug design heavily relies on calculating this binding strength. 
For example, Virtual Screening (VS) tries to find the best chemical compounds which can regulate the function of a protein; in drug discovery this is often a disease-related target protein. 
VS achieves this goal by assigning a score to each binding ligand, indicating how strong it binds to the target protein. 
To get the overall picture of how binding affinity prediction enables VS, the reader is referred to \cite{c2014virtual, lionta2014structure, wingert2018improving}. 

Current approaches for quantifying the binding affinity can be categorized as physics-based, empirical, knowledge-based and descriptor-based scoring functions \cite{liu2015classification}. 
In spite of their merits, the conventional techniques assume a predetermined functional form which is additive. 
Furthermore, they need domain knowledge to extract features and formulate the scoring functions. 
For example, the semi-empirical force field in AutoDock \cite{morris2009autodock4} and empirical scoring function in X-Score \cite{wang2003comparative} belong to this category. 
As instance, X-Score takes average of three scoring functions HPScore, HMScore, and HSScore, differing by the terms which describe the hydrophobic interactions. 
Each of these scoring functions comes in the form of a linear combination of the terms \cite{wang2002further}. 

Only in the past decade the machine learning algorithms have been used to score the protein-ligand binding strength in a data-driven manner. 
Ballester and coworkers gathered data about the frequency of occurrence of each possible protein-ligand atom pair up to a distance threshold. 
By binning these counts, they trained their Random Forest model implicitly on short-range, middle-range, and long-range interactions. 
The RF-Score model is still among the top performer models in this field \cite{ballester2014does}. 
However, such models also heavily rely on biological feature engineering to extract descriptors or fingerprints; it is still based on expert knowledge and therefore biased. 

Deep learning models, which belong to descriptor-based category, aim to minimize the bias due to expert knowledge.
To describe the interactions between a protein and its ligand, atom-centered and grid-based methods are the most widely used techniques. 
Schietgat \textit{et al.} developed a 3D neighborhood kernel, which took the atom spatial distances into consideration, to describe the structure of proteins and ligands \cite{schietgat2015predicting}. 
More recently Gomes and coworkers \cite{gomes2017atomic} represented the structures of proteins and ligands as a combination of neighbor lists and atom types, to later be used in a deep network. 
In their case, they described the \textit{whole} protein and ligand structures using their atom-centered scheme. 

By contrast, grid-based approach usually limits the representation of protein and ligand interactions to a grid box defined around a protein's binding site, and different atom information is encoded in different channels of the 3D grid. 
Wallach \textit{et al.}\cite{wallach2015atomnet} and Ragoza \textit{et al.} \cite{ragoza2017protein} developed CNN scoring functions to classify compound poses as binders or non-binders. 
Jim\'enez \textit{et al.} \cite{jimenez2018k} and Marta \textit{et al.} \cite{stepniewska2018development} designed similar deep learning approachs to predict the binding affinity, based on the rasterized protein-ligand structure input.

In addition to modeling approach, data reliability is a major issue for binding affinity prediction.
Although a few thousands of labeled protein-ligand complexes are available, their binding affinity are reported as different experimental measures, including $K_{d}$, $K_{a}$, $K_{i}$, and $IC_{50}$, in decreasing order of reliability for the purpose of binding affinity prediction. 
If we indiscriminately feed all the data to the machine learning model, it will potentially introduce label noise or even incorrect labels different from ground truth. 
The machine learning model may then suffer from the inaccurate supervision. 
    
Our goal in this paper is twofold. 
First, we aim to develop an end-to-end solution for binding affinity accurate prediction which 1) gets as input the 3D structural data for the protein and ligand, 2) requires minimum feature engineering, and 3) achieves state-of-the-art prediction performance.
Second, we aim to systematically analyze the publicly available protein-ligand binding affinity data and propose a new benchmark for more reliable model learning. 

More recently, deep learning has exhibited its powerful superiority in a broad range of bioinformatics tasks, such as gene mutations impact prediction\cite{sundaram2018predicting}, protein folding\cite{li2018foldingzero} and drug discovery\cite{segler2018planning}.
By stacking many well-designed neural network layers, it is capable of extracting useful features from raw data form and approximating highly complex functions \cite{lecun2015deep}. 
Many advanced deep learning algorithms are developed based on convolutional neural networks (CNNs). 
The impressive performance of CNNs is mainly because they can effectively take advantage of spatially-local correlation in the data. 
Similarly, protein-ligand 3D structure naturally has such characteristics; biochemical interactions between atoms occur locally. 
CNNs hopefully can hierarchically compose such local spatial interactions into abstract high-dimensional global features contributing to the binding score.

In this paper, we propose the framework DeepAtom to accurately predict the protein-ligand binding affinity. 
A complex of protein-ligand is first rasterized into a 3D grid box, which is centralized on the ligand center. 
Each voxel has several input channels, embedding the different raw information of atoms located around the voxel. 
A light-weight 3D-CNN model is then developed to hierarchically extract useful atom interaction features supervised by the binding affinity score.
As a data-driven approach, it effectively avoids \textit{a priori} functional form assumptions.
More importantly, our efficient architecture design significantly improves the model representational and generalization capacity even trained with the limited available complex data.

We present comprehensive experiments on the standard benchmark test set, called PDBbind v.2016 \textit{core} set \cite{wang2005the} and an additional test set, called Astex Diverse Set \cite{hartshorn2007diverse}. 
Randomly initialized and evaluated for 5 times, DeepAtom consistently outperforms all the state-of-the-art published models. 
In order to further improve the model performance, we also critically study the publicly available complex data and propose a new benchmark dataset.
It further improves the DeepAtom performance, with potential benefits to the future research in the binding affinity scoring field.

\section{Materials and Methods}
\subsection{Input Featurization and Processing}
\subsubsection{Protein-ligand Complex} 
The standard datasets, such as PDBbind and Binding MOAD, include the structures of protein and ligand in their bound form, a.k.a. their complex, deposited in a single PDB file. 
Based on experimental techniques, the strength of ligand binding to protein has been determined for each structure. This binding affinity data is used as the ground truth labels for these protein-ligand complexes. 

\begin{figure}  %h=horizontal, v=vertical, t=top
    \centering
    \includegraphics[scale=0.32]{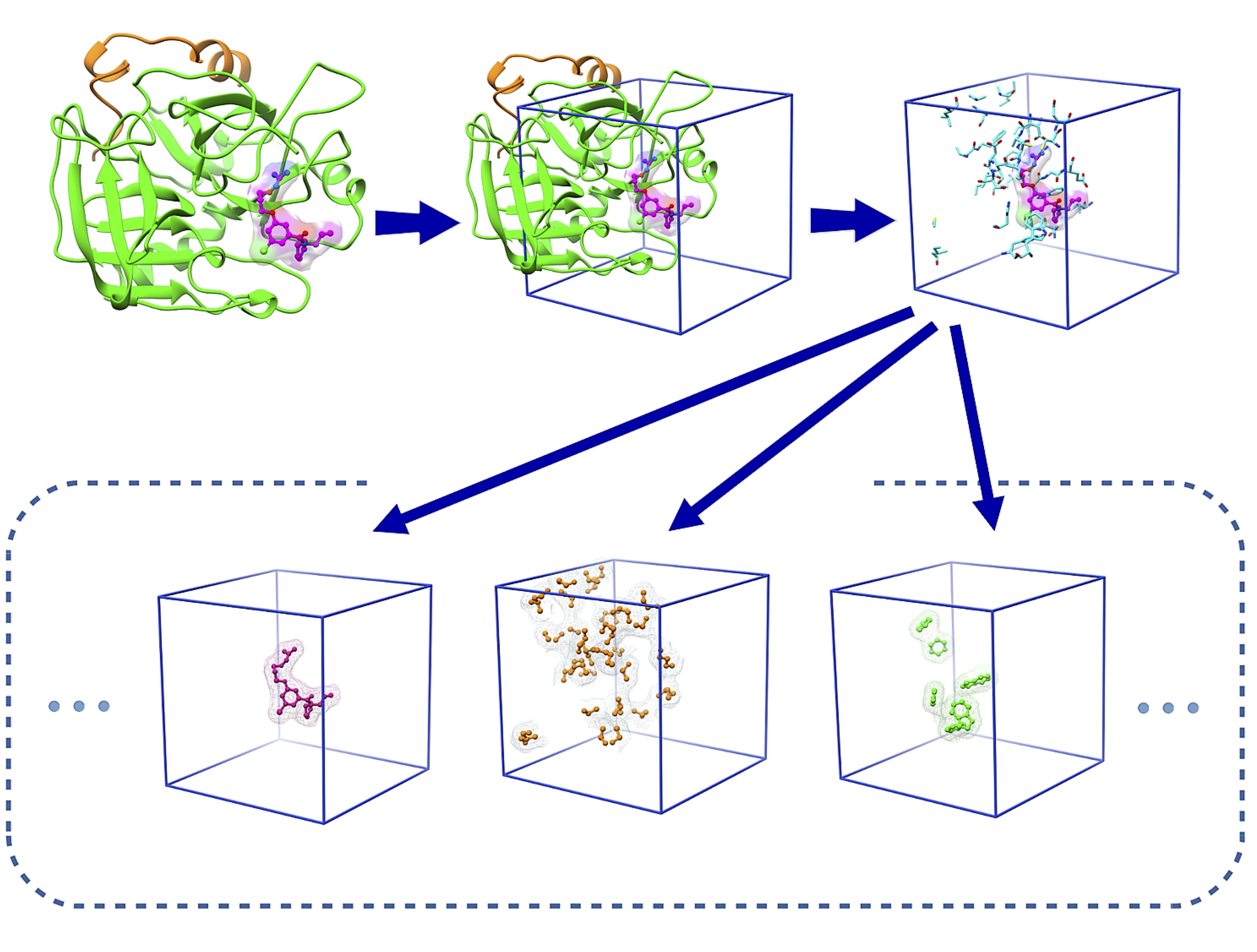}  %scale of figure % width=\linewidth
    % \vspace{-1em}
    \setlength\belowcaptionskip{-1.7\baselineskip}
    \caption{\textbf{Local box featurization (3D data representation).} The grid box encompasses the area around the binding site, centered on the ligand. Each channel includes only a specific feature, e.g. from left to right, the three channels shown are the excluded volume channel for the ligand as well as the hydrophobic and aromatic channels for protein. Each sample is described in terms of 24 channels in total.
}
    \label{fig:Featurization}
    \vspace{1.0em}
\end{figure}

\subsubsection{Grid Size \& Resolution} 
We calculate the distribution of end-to-end distances for all ligands in the PDBbind v.2016 \textit{refined} and \textit{core} datasets. 
This gives us clues to define a box size of \num{32} \AA, which is the same as the end-to-end distance for the longest ligand in these two datasets, so there is no need to filter out any. 
The distribution of ligand lengths in these two subsets is illustrated in Fig.~\ref{fig:merge}a.

The van der Waals radius of the 9 major heavy atoms (C, N, O, P, S, F, Cl, Br, I) used in our study is greater than 1.4 \AA. 
As a simplified view, an atom's $r_{vdw}$ can be assumed as a measure of its size; it is defined as half of the internuclear separation of two non-bonded atoms of the same element on their closest possible approach. 
A grid resolution larger than $2\times r_{vdw}$ cannot differentiate two atoms from each other. On the other hand, a finer resolution brings about much higher computational cost. 
As a trade-off between accuracy and efficiency, we set the grid resolution as 1.0 \AA.

\subsubsection{Features / Atom Types} 
\label{sec:atom_types}
The 11 Arpeggio atom types are based on the potential interactions each atom may get involved in \cite{jubb2017arpeggio}; they include features such as Hydrogen bond acceptor and donor, positive or negative, hydrophobic, and aromatic atom types. 
The protein and ligand atoms are described by 11 Arpeggio atom types and an excluded volume feature, where discrimination is made between protein and ligand atoms. 
This resulted in $(11+1)\times2=24$ features.

\subsubsection{Occupancy Type} This hyper-parameter defines how each atom impacts its surrounding environment. 
In our work, each atom can affect its neighbor voxels up to double of its van der Waals radius $r_{vdw}$ through a pair correlation potential. 
We use the Atom-to-voxel PCMax algorithm, described in \cite{jimenez2017deepsite}, where each  atom makes a continuous contribution $n(r)$ to its neighbor voxels as defined by Eq.~\ref{eq:pc_max}.
At the center of a voxel, only the maximum effect from contributing atoms is kept. 
\begin{equation}
\setlength{\abovedisplayskip}{2pt}
    n(r)=1-exp(-(\frac{r\textsubscript{vdw}}{r})^{12})
    \label{eq:pc_max}
\setlength{\belowdisplayskip}{2pt}
\end{equation}

\subsection{Network Architecture}
\begin{figure*}[htb]  %h=horizontal, v=vertical, t=top
    \centering
    \includegraphics[scale=0.5]{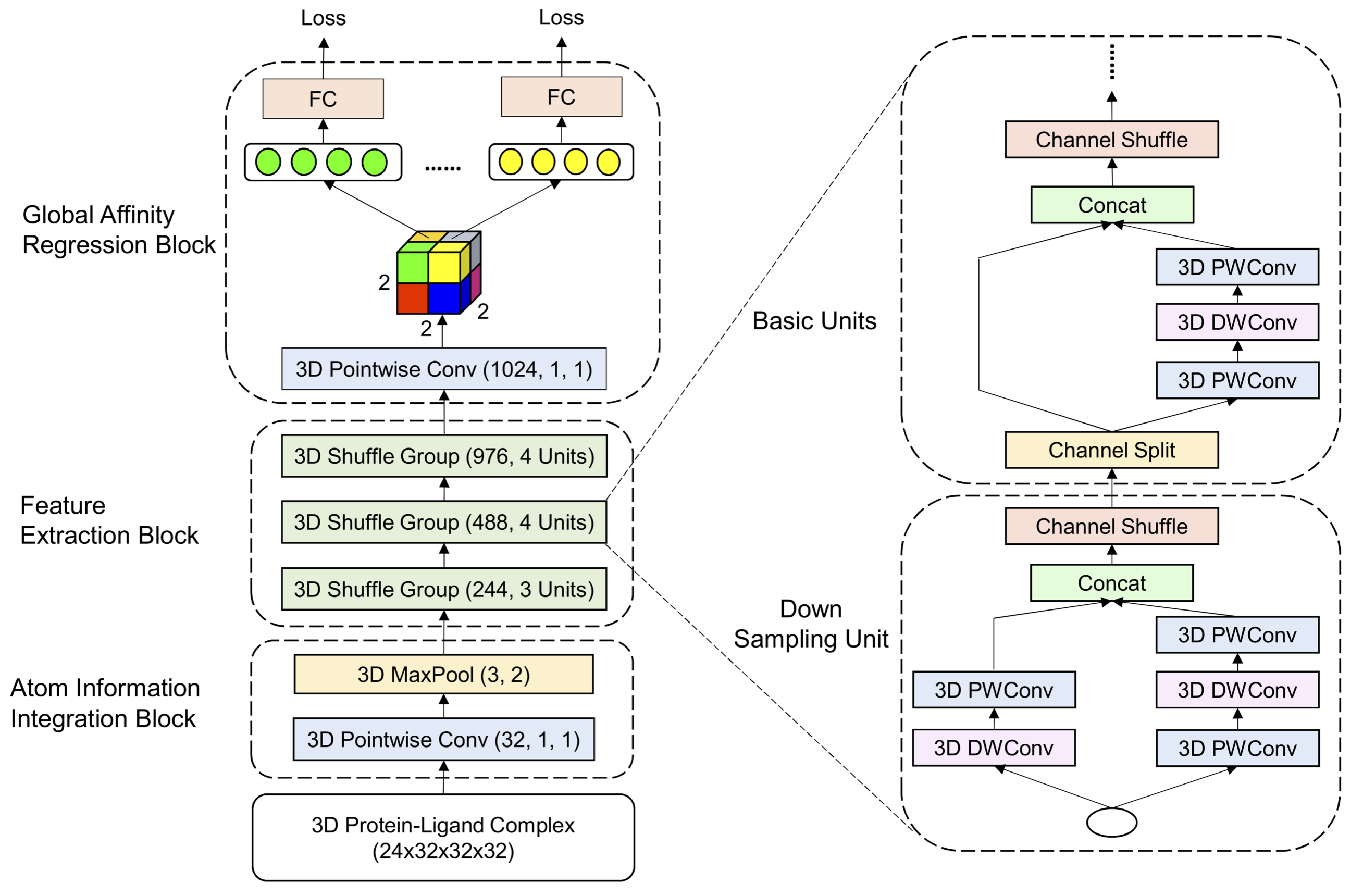}  %scale of figure % width=\linewidth
    \setlength{\belowdisplayskip}{3pt}
    \caption{\textbf{Network architecture.} Each Conv layer is specified by its number of channels, kernel size and stride. The 3D MaxPool layer has kernel size 3 and stride 2. For the 3D Shuffle Groups, the numbers in parentheses denote the number of output channels and repeat time of the unit. Only the first unit has down sampling layer, where the DWConv layer has kernel size 3 and stride 2. In the remaining units, DWConv with kernel size 3 and stride 1, as well as PWConv with kernel size 1 and stride 1 are utilized. Eight losses are calculated based on the shared weights FC layer output. Two dropout layers are appended before the last 3D Pointwise Conv and FC layers respectively.
}
    \label{fig:Network}
    \vspace{-1.2em}
\end{figure*}

To extract the atomic interaction information from the voxelized protein-ligand complex data, a straightforward approach is to extend 2D-CNNs to 3D by using 3D convolution kernels. 
One channel of the output feature map at the location $(i, j, k)$ is computed by the standard convolution as follows:
\begin{equation}
    \operatorname{Conv}(W, h)_{(i, j, k)}=\sum_{s, t, r, m}^{S, T, R, M} W_{(s, t, r, m)} \cdot h_{(i+s, j+t, k+r, m)}
    \label{Conv}
\end{equation}
where $h$ represents the input with $M$ channels, $W_{(s,t,r,m)} \in \mathbb{R}^{S\times T\times R \times M}$ represents one filter weight, and $S, T, R$ are side length of the filter. 
However, 3D convolution itself will massively inflate the amount of trainable network parameters, due to the increase in the input and kernel dimensions. 
Specifically, if $N$ is the number of the output channels, one standard convolution layer will introduce $S \cdot T \cdot R \cdot M \cdot N$ parameters.
% \begin{equation}
%      \cdot F \cdot F \cdot F
% \end{equation}
More importantly, in order to improve the learning ability and achieve higher prediction accuracy, a general trend has been to make the model deeper and more complicated\cite{simonyan2014very},\cite{he2016deep,huang2017densely}. 
By contrast, for the affinity prediction problem, only a few thousands of protein-ligand complexes with experimentally determined binding affinity are available. 
This issue discourages the use of network architectures with too many trainable parameters, because overfitting may occur when the network has high complexity whereas a relatively small data set is available for training. 
Indeed, another \num{3}D CNN-based affinity prediction work \cite{ragoza2017protein} also encounters the overfitting issue. 
After empirically optimizing the model depth and width, they ultimately reduce the network to only three convolutional layers.
Similarly, Pafnucy \cite{stepniewska2018development} is developed as a \num{3}D CNN model with three convolutional and three dense layers.

Our model is inspired by the light-weight network architecture, and aims to achieve the best trade-off between the prediction accuracy and the model complexity in terms of learnable parameters.
A series of related network structure have been recently proposed, such as Xception~\cite{chollet2017xception}, MobileNet v1~\cite{howard2017mobilenets} and v2~\cite{sandler2018mobilenetv2}, ShuffleNet v1~\cite{zhang2018shufflenet} and v2~\cite{ma2018shufflenet}, and CondenseNet~\cite{Huang_2018_CVPR}. 
Based on the practical guidelines for efficient CNN architecture design and the corresponding ShuffleNet units described in \cite{ma2018shufflenet}, we propose a novel light-weight \num{3}D CNN mdoel, which can be effectively trained with deeper layers by the limited training samples. 
It improves the prediction performance by a large margin, but does not significantly increase model complexity.

Specifically, as shown in Fig.~\ref{fig:Network} our model consists of three building blocks, namely atom information integration block, stacked feature extraction block, and global affinity regression block.

In the atom information integration block, a pointwise (PW, $1\times1\times1$) convolution layer defined in Eq.~\ref{eq:PW} with non-linear activation function is first utilized to fuse the atom information across different channels. 
\begin{equation}
\setlength{\abovedisplayskip}{2pt}
    \operatorname{PWConv}(W, h)_{(i, j, k)}=\sum_{m}^{M} W_{m} \cdot h_{(i, j, k, m)}
    \label{eq:PW}
\setlength{\belowdisplayskip}{2pt}
\end{equation}
This cascaded cross-channel parametric pooling structure brings about an improvement compared with the empirical scoring functions. 
For instance, AutoDock software's scoring function is composed of a linear combination of interaction types, such as Hydrogen bonding and electrostatic interactions \cite{morris2009autodock4}. The pointwise convolution layer in our model is followed by a 3D max pooling layer to increase the translational invariance of the network and reduce the input dimension. 
The output of this block has the grid size of $16\times16\times16$. 

The feature extraction block consists of multiple consecutive 3D shuffle units, and according to the number of channels in their outputs, they are categorized into three groups. 
At the beginning of the unit, a channel split operator equally splits the input of feature channels into two branches. 
One branch data is sequentially processed by a pointwise convolution, a $3\times3\times3$ depthwise (DW) convolution and an additional pointwise convolution. 
All three layers have the same number of input and output channels $N$. 
Depthwise convolutional layer performs the spatial convolution independently over every channel of an input:
\begin{equation}
\setlength{\abovedisplayskip}{3pt}
    \operatorname{DWConv}(W, h)_{(i, j, k)}=\sum_{s, t, r}^{S, T, R} W_{(s, t, r)} \odot h_{(i+s, j+t, k+r)}
\setlength{\belowdisplayskip}{3pt}
\end{equation}
where $\odot$ denotes the element-wise product.
Although depthwise convolution does not combine different input channels, the two neighbor regular pointwise convolution effectively fuse the information across the channels.
The other branch is kept as identity until it is concatenated with the output from the first branch. 
This identity branch can be regarded as an effective feature reuse design, which strengthens feature propagation and reduces the number of parameters. 
Within a basic unit, the depthwise and pointwise convolutions respectively introduce $S \cdot T \cdot R \cdot \frac{N}{2}$ and $\frac{N}{2} \cdot \frac{N}{2}$ parameters. 
Therefore, using a basic unit to replace the standard convolution, we obtain the parameter reduction as follows:
\begin{equation}
    \frac{S \cdot T \cdot R \cdot \frac{N}{2} + 2 \cdot \frac{N}{2} \cdot \frac{N}{2}} 
    {S \cdot T \cdot R \cdot N \cdot N } 
    = \frac{1}{2} (\frac{1}{N} + \frac{1}{S \cdot T \cdot R})
    % \frac{1}{2 \cdot N} + \frac{1}{2 \cdot S \cdot T \cdot R} 
    \label{eq:Reduce}
\end{equation}
DeepAtom uses $3\times3\times3$ depthwise convolutions and the number of channels are set as $244, 488, 976$. 
Therefore, with the efficient model design, we can easily obtain more than 20 times parameters reduction, which enable us to stack deeper layer to improve the model learning capacity.

At the end of the units, the channel shuffle operation is applied to enable the information flow across the two branches.
Particularly, the channel shuffle operation first divides the feature map in each branch into several subgroups, then mixes the branches with different subgroups. 
When the spatial down sampling is applied, the channel split operator in the shuffle unit is removed, and the number of output channels is doubled. 
In each group, only the first shuffle unit has the down sampling layer, and the remaining units keep the input dimension.

After stacking three shuffle groups, the original 3D input data is down sampled to a $1024\times2\times2\times2$ 4D tensor (3 grids along x, y, z axes and \num{1024} channels). 
The global affinity regression block first slices the tensor into $2\times2\times2=8$ vectors with dimension \num{1024}. 
Based on the prior shuffle groups, the receptive field of each vector covers the entire raw volume, so we set up the affinity prediction task for each vector.
A shared weights fully connected (FC) layer consumes each vector to construct regression loss, and it enables us to train the top layers more thoroughly and further avoid overfitting. 
In testing phase, outputs from the multiple hidden vectors are averaged right before the FC layer to stabilize the prediction.

In the architecture, we adopt the leaky rectified linear unit as the activation function. A batch normalization layer is appended after each convolution operation to speed up the training. The mean squared error is set as our affinity regression loss for model learning.

\subsubsection{Training} The model is updated by Adam algorithm with default parameters for momentum scheduling ($\beta_1$ = \num{0.9}, $\beta_2$ =\num{0.999}). Training the model from scratch, we set the initial learning rate as 0.001 and the weight decay to \num{4e-5}. 
Our model is implemented using PyTorch (version \num{0.4}). With batch size of \num{256}, the model is trained for around \num{60} epochs on \num{2} Nvidia P\num{40} GPU cards. 

\subsubsection{Data Augmentation} 
The publicly available biological datasets contain only thousands of complexes with reliable experimental binding affinity value. 
Directly training on these insufficient samples easily makes the deep learning model suffer from the overfitting problem.
Data augmentation is proved as an effective approach to enhance the deep learning model performance. 
In our experiments, each of the original samples gets randomly translated and rotated. 
Enabling such transformations significantly improves the training and model capacity. 
In order to reduce the variance, the augmented samples of each protein-ligand complex are averaged during the prediction phase.

\subsection{Dataset Preparation}
\subsubsection{PDBbind Dataset}
\label{sec:first trainset}
PDBbind is the standard dataset for developing models to predict the binding affinity of protein-ligand complexes \cite{wang2005the}; it has three subsets, namely \textit{core}, \textit{refined}, and \textit{general}. 
The \textit{general} subset includes complexes with relatively lower quality; it contains lower resolution structures, and the experimental affinities for some structures are reported as $IC_{50}$ values. 
The \textit{refined} dataset is a higher quality subset of the \textit{general} dataset; it includes complex structures with better resolution and excludes any complex with $IC_{50}$ data only. 
In total, PDBbind v.\num{2016} \textit{refined} dataset includes \num{4057} complexes.
The \textit{core} dataset is a subset of \textit{refined} data, clustered with a threshold of $90\%$ sequence similarity; five representative complexes are picked for each of the \num{58} protein family clusters in order to cover the affinity range better. 
This results in \num{290} complexes in the \textit{core} subset, which serves as the standard test set to evaluate the scoring approaches.  
We further split the rest of \num{3767} non-overlapping complexes between the \textit{refined} and \textit{core} into two disjoint subsets: $(\romannumeral1)$ $10\%$ of complexes (\num{377}) are randomly selected and used as the validation set, $(\romannumeral2)$ the rest ($3390$ complexes) are used for training, which is named as \enquote{training set-1}.

\subsubsection{Proposed Benchmark Training Set}
\label{sec:benchmark dataset}
In order to compile an improved benchmark dataset, we use PDBbind data as well as a complementary source of protein-ligand binding affinity data, namely Binding MOAD \cite{hu2005binding,ahmed2015recent}.
In order to incorporate the recently updated complexes, we start from PDBbind v.\num{2018} dataset, and extract all complexes with either $K_{d}$, $K_{a}$, or $K_{i}$ data from \textit{general} and \textit{refined} subsets.
It is worth noting that we exclude the complexes shared with the \textit{core} subset to prevent the data leakage. 
We follow the same steps with Binding MOAD data. 
A few filtration steps are also necessary: first if a complex has reported $K_{d}/K_{a}$ data in one database and $K_{i}$ in the other, we keep the $K_{d}/K_{a}$ data only. 
Second, complexes with a peptide as their ligand are discarded. 
We do not filter the complexes based on their structure resolution nor perform any clustering on them in terms of protein sequence or structure; clustering is typically done to later reduce the dataset into representative samples. 
The limited availability of the experimental affinity data discourages further removal of samples, although the dataset is biased towards some structures, e.g. the congeneric series. 

In total, the final benchmark dataset contains \num{10383} complexes.
Please note that in contrast to NMR structures which contain multiple 3D models, a PDB file from X-ray crystallography contains a single 3D structure only. 
Our proposed benchmark dataset includes almost exclusively X-ray structures, with only one structure existing in each PDB file. Merely 63 complexes come from NMR experiments. 
We get only the first model from these PDB files.

Compared with the \textit{refined} subset of PDBbind, this dataset almost doubles the number of samples with $K_{d}/K_{a}/K_{i}$ data and is expected to improve the performance of binding affinity scoring techniques. 
The full list of the proposed benchmark dataset for model training is provided in the \href{https://tinyurl.com/y6roltyd}{Supplementary Table\footnote{\href{https://github.com/YanjunLi-CS/DeepAtom_SupplementaryMaterials}{https://github.com/YanjunLi-CS/DeepAtom\_SupplementaryMaterials}}}. 
The $pK_{d}/pK_{a}/pK_{i}$ value for each complex is reported to make it easier for other researchers to use the proposed dataset. 
The binding score of the complexes in this dataset ranges from $-0.15$ to $15.22$ in $pk$ units, and the score distribution is shown in \href{https://tinyurl.com/yxdakv3c}{Supplementary Fig. S1}.

To train the scoring approaches, we split the proposed benchmark dataset into two subsets:
$(\romannumeral1)$ we randomly select $1000$ samples from non-overlapping complexes between PDBbind v.\num{2016} \textit{refined} and \textit{core} sets.
$(\romannumeral2)$ the rest ($9383$ complexes) are for training, named as \enquote{training set-2}.

\subsubsection{Astex Diverse Set} This dataset was developed in 2007. It includes 85 protein-ligand complexes filtered to be of interest specifically to pharmaceutical and agrochemical industries \cite{hartshorn2007diverse}. Among these 85 complexes, 64 of them include binding affinity data. 

\subsection{Other Methods for Comparison}
In Section \ref{section:results}, we compare DeepAtom with three state-of-the-art and open-source scoring approaches: Pafnucy model \cite{stepniewska2018development}, RF-Score \cite{ballester2014does}, and X-Score \cite{wang2003comparative}.
For Pafnucy and RF-Score, we use the open-source codes provided by the authors, and use their suggested hyper-parameters to re-train models on the same datasets as DeepAtom.
For X-Score, we take the results from paper \cite{jimenez2018k}, where the authors used the publicly available binaries to make predictions on the same PDBbind v.\num{2016} \textit{core} set.

\begin{figure*}[htb]  %h=horizontal, v=vertical, t=top
    \centering
    \includegraphics[scale=0.77]{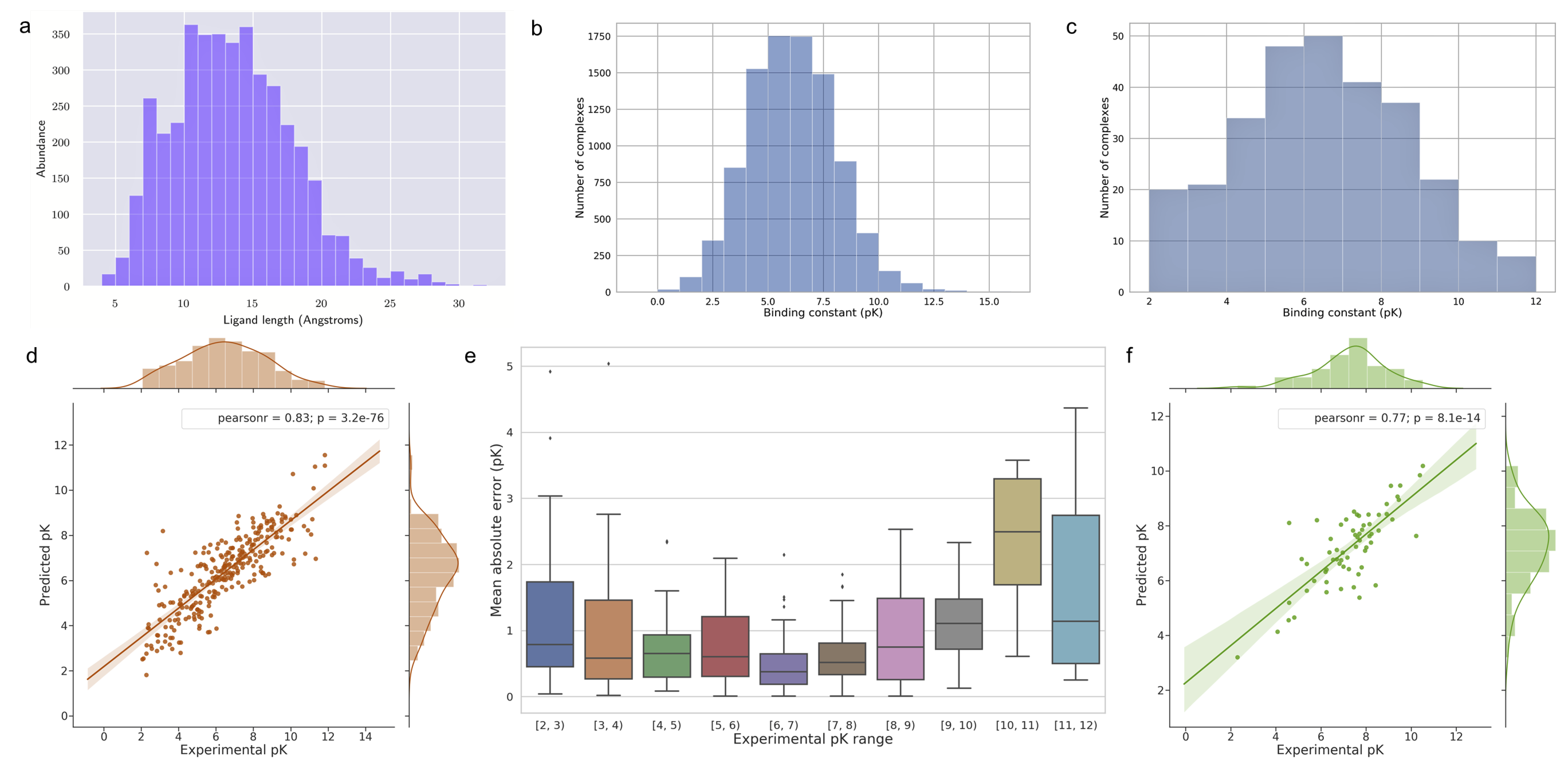}  %scale of figure % width=\linewidth
    \vspace{-0.5em}
    \setlength{\belowdisplayskip}{0.5pt}
    \caption{\textbf{a}\label{fig:3}. Ligand length distribution in the PDBbind v.2016 \textit{refined} and \textit{core} sets. \textbf{b}. Binding data distribution of the training set in our proposed benchmark. \textbf{c}. Binding data distribution of the \textit{core} set. \textbf{d}. DeepAtom prediction results for the \textit{core} set. \textbf{e}. The distribution of MAE between DeepAtom prediction and target complexes with different binding ranges. \textbf{f}. DeepAtom prediction results for the Astex Diverse Set.}
    \label{fig:merge}
    \vspace{-1em}
\end{figure*}

\section{Results \& Discussion}
\label{section:results}
In this section, we describe the training and benchmarking protein-ligand complex data for DeepAtom. 
The evaluation details are presented along with discussion of the results. 

\subsection{Evaluation Metrics}
To comprehensively evaluate the model performance, we use Root Mean Square Error (RMSE) and Mean Absolute Error (MAE) to measure the prediction error, and use Pearson correlation coefficient (R) and standard deviation (SD) in regression to measure the linear correlation between prediction and the experimental value.
The SD is defined in Eq.~\ref{eq:sd}.
\begin{equation}
    SD=\sqrt{\frac{\sum_{i}^{n}\left[y_{i}-\left(a+b x_{i}\right)\right]^{2}}{n-1}}
    \label{eq:sd}
\end{equation}
where $x_i$ and $y_i$ respectively represent the predicted and experimental value of the $i$th complex, and $a$ and $b$ are the intercept and the slope of the regression line, respectively.

\begin{table}
\caption{Results on PDBbind v.\num{2016} \textit{core} set with \enquote{training set-1}. 
In each table cell, mean value over five runs is reported as well as the standard deviation in parentheses.}
\label{tb:Core_tr_PDBbind}
\renewcommand{\arraystretch}{1.5}
\centering
\scalebox{0.9}{
\begin{tabular}{ccccc}
\hline
 & RMSE & MAE & SD & R \\ \hline
DeepAtom & \textbf{1.318 (0.212)} & \textbf{1.039 (0.016)} & \textbf{1.286 (0.015)} & \textbf{0.807 (0.005)} \\ \hline
RF-Score & 1.403 (0.002) & 1.134 (0.003) & 1.293 (0.002) & 0.803 (0.001) \\ \hline
Pafnucy & 1.553 (0.031) & 1.261 (0.027) & 1.521 (0.037) & 0.722 (0.017) \\ \hline
\end{tabular}
}
\vspace{-2em}
\end{table}

\subsection{Model Comparison with \enquote{Training Set-1}}
We first train DeepAtom, RF-Score and Pafnucy on the \enquote{training set-1} with \num{3390} complexes described in Section \ref{sec:first trainset}, and evaluate them on the PDBbind v.\num{2016} \textit{core} set, which is unseen to the model during its training and validation.
Each approach is randomly initialized and evaluated for \num{5} times. 
The mean and the standard deviation (in the parentheses) of the four evaluation metrics are presented in Table \ref{tb:Core_tr_PDBbind} for testing, and \href{https://tinyurl.com/yxdakv3c}{Supplemental Table S1} for validation. 
Learning with the very limited samples, DeepAtom outperforms the similar 3D CNN-based Pafnucy by a large margin, which demonstrates that our light-weight architecture design enables effective training with the deep layers and significantly improves its learning and generalization capacity.
% and can effectively reduce the risk of overfitting problem.
On the other hand, DeepAtom achieves the comparable performance with the conventional machine learning method RF-Score, although as a practical guideline, training a supervised deep learning model generally requires larger datasets.
It suggested that our model has greater potential to provide more accurate prediction, given enough training data.

\subsection{Model Comparison with \enquote{Training Set-2}}
\begin{figure}[t]
  \begin{subfigure}[t]{0.24\textwidth}
    \includegraphics[width=\textwidth]{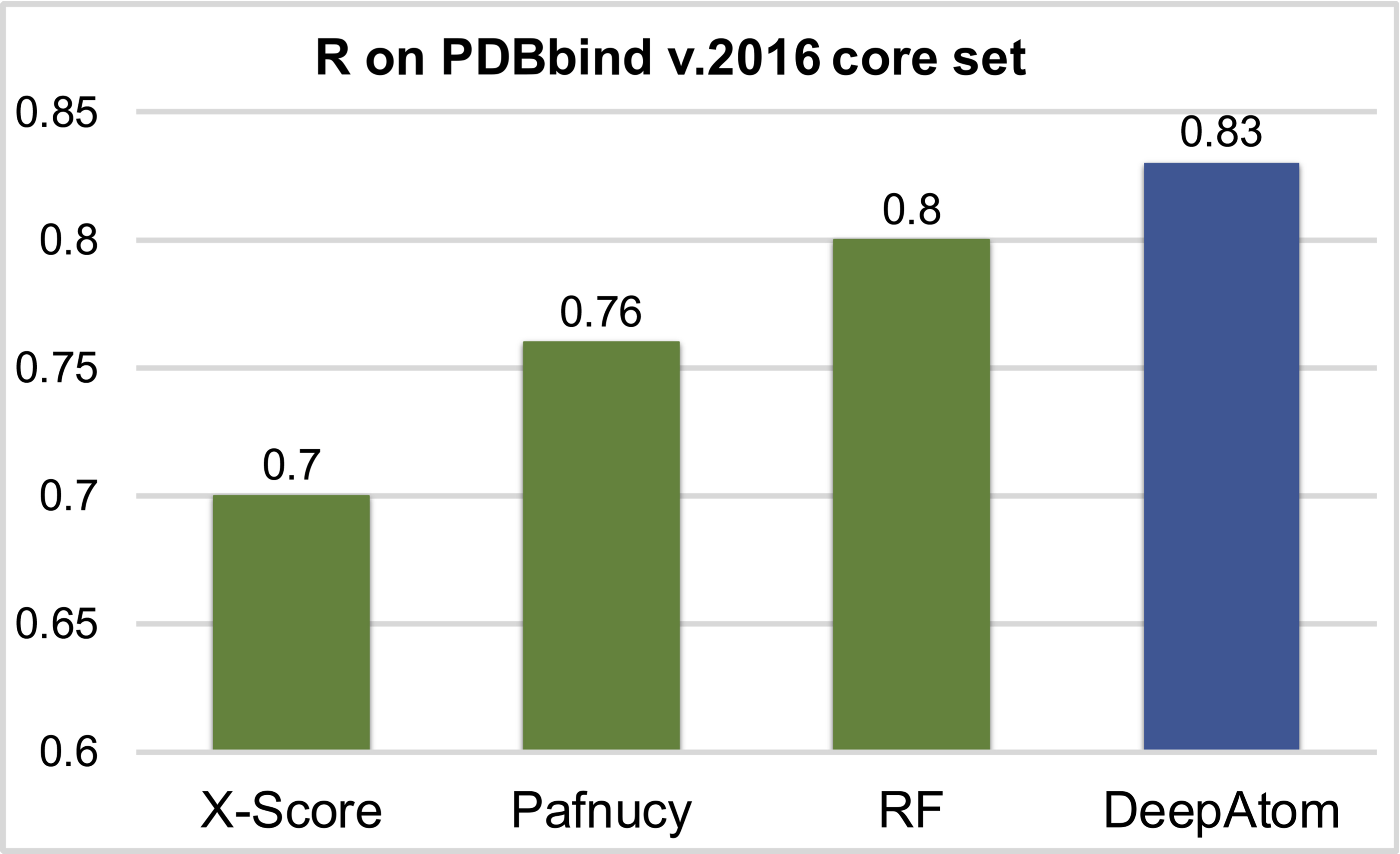}
    \caption{Comparison of R}
    \label{fig:R}
  \end{subfigure}
  \begin{subfigure}[t]{0.24\textwidth}
    \includegraphics[width=\textwidth]{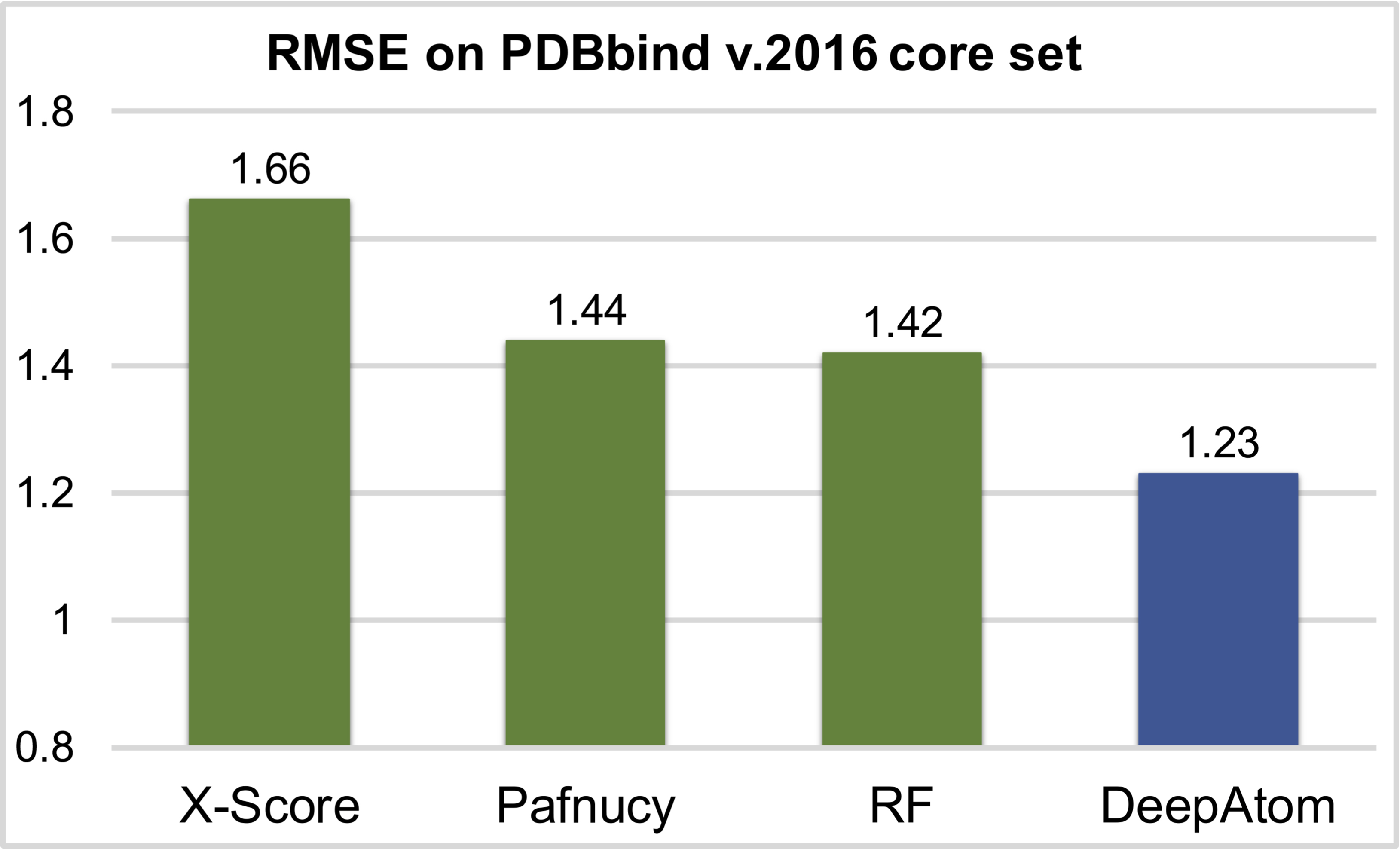}
    \caption{Comparison of RMSE}\label{fig:RMSE}
  \end{subfigure}
  \caption{Comparison of scoring methods on PDBbind v.\num{2016} \textit{core} set.}
  \label{fig:compare_other_score}
\vspace{-0.5em}
\end{figure}

% \vspace{-0.5em}
Next, we use our proposed \enquote{training set-2} to re-train DeepAtom, RF-Score and Pafnucy models, and also evaluate them on the PDBbind v.2016 \textit{core} set.
Similarly,  \num{5} different runs are conducted for each scoring method to stabilize the results.
Fig.~\ref{fig:compare_other_score} shows the comparison results in terms of mean value of the R and RMSE, also including the X-Score prediction results. 
Table \ref{tb:Core_tr_Benchmark} presents the detailed results of the top three methods.
As expected, DeepAtom outperforms all the other approaches across all four measurements by a large margin. 
It obtains the best Pearson correlation coefficient of \num{0.83} and RMSE of \num{1.23} in $pK$ units, compared with RF-Score results: $R=0.80$ and $RMSE=1.42$, and Pafnucy results: $R=0.76$ and $RMSE=1.44$.
To the best of our knowledge, DeepAtom achieves the state-of-the-art performance on this well-known benchmark.
The corresponding validation results are shown in the \href{https://tinyurl.com/yxdakv3c}{Supplementary Table S2}.

Fig.~\ref{fig:merge}d shows the correlation between the prediction results of one DeepAtom model and the experimental binding affinity data. 
DeepAtom gives the highly correlated prediction on the PDBbind v.2016 \textit{core} set. 
To further investigate the model performance on different ranges of the binding data, we visualize the binding affinity distribution of the training set (Fig.~\ref{fig:merge}b) in the proposed benchmark, PDBbind v.2016 \textit{core} set (Fig.~\ref{fig:merge}c) and the corresponding DeepAtom RMSE value within each $pK$ unit range (Fig.~\ref{fig:merge}e). 
From Fig.~\ref{fig:merge}b and \ref{fig:merge}c, we observe that the binding scores of our training samples are intensively located in the middle range (from \num{3} to \num{9} $pK$ units) which is highly similar to the \textit{core} set. 
Fig.~\ref{fig:merge}e shows that DeepAtom obtains better prediction results with lower MAE values in this middle range, compared to the less frequent binding scores. 
For a data-driven approach, the distribution of training data plays a crucial role in its performance.
Because the number of training samples falling in the middle range is much larger than the samples with marginal affinity values, DeepAtom naturally performs better for the complexes in the middle range during the testing stage.
It also suggests that diversifying the training samples is promising for DeepAtom to provide more accurate and reliable predictions.

\begin{table}
\caption{Results on PDBbind v.\num{2016} \textit{core} with \enquote{training set-2}.}
\label{tb:Core_tr_Benchmark}
\renewcommand{\arraystretch}{1.5}
\scalebox{0.9}{
\begin{tabular}{ccccc}
\hline
 & RMSE & MAE & SD & R \\ \hline
DeepAtom & \textbf{1.232 (0.011)} & \textbf{0.904 (0.019)} & \textbf{1.222 (0.011)} & \textbf{0.831 (0.003)} \\ \hline
RF-Score & 1.419 (0.002) & 1.124 (0.001) & 1.304 (0.002) & \multicolumn{1}{l}{0.801 (0.000)} \\ \hline
Pafnucy & 1.443 (0.021) & 1.164 (0.019) & 1.424 (0.022) & 0.761 (0.008) \\ \hline
\end{tabular}
}
% \vspace{-0.5em}
\end{table}

\begin{table}
\caption{Results on Astex Diverse Set with \enquote{training set-2}.}
\label{tb:astex}
\renewcommand{\arraystretch}{1.5}
\scalebox{0.89}{
\begin{tabular}{ccccc}
\hline
 & RMSE & MAE & SD & R \\ \hline
DeepAtom & \textbf{1.027 (0.061)} & \textbf{0.714 (0.033)} & \textbf{1.003 (0.042)} & \textbf{0.768 (0.022)} \\ \hline
RF-Score & 1.144 (0.006) & 0.891 (0.010) & 1.103 (0.004) & \multicolumn{1}{l}{0.710 (0.003)} \\ \hline
Pafnucy & 1.374 (0.057) & 1.110 (0.065) & 1.288 (0.039) & 0.569 (0.04) \\ \hline
\end{tabular}
}
\vspace{-1.5em}
\end{table}

We also compare DeepAtom with RF-Score and Pafnucy on the independent Astex Diverse Set.
Table \ref{tb:astex} shows that DeepAtom again significantly outperforms the others over all the measurements.
The prediction results from one DeepAtom model are illustrated in Fig.~\ref{fig:merge}f.

\subsection{Evaluation of the Proposed Benchmark Dataset}
Comparison of Tables \ref{tb:Core_tr_PDBbind} and \ref{tb:Core_tr_Benchmark} reveals that training on our proposed benchmark dataset results in a significant improvement of the model performances, especially for the deep leaning based approaches.
For example, DeepAtom increases the R from \num{0.81} to \num{0.83} and decreases the RMSE from \num{1.32} to \num{1.23}.
The difference between the two tables confirms the effectiveness of our proposed new benchmark dataset, where the model trained on our new dataset provides more accurate predictions.   
Although the dataset contains some complexes with low resolution structure, such low-quality data does not introduce obvious label noise. 
On the contrary, this extended dataset provides more reliable complex data which can effectively improve the generalization power of binding affinity prediction algorithms.

It is worth noting that our proposed benchmark dataset extends the standard \textit{refined} set by including the complexes from PDBbind \textit{general} subset and Binding MOAD database only when the experimental affinity data is either $K_{d}$, $K_{a}$, or $K_{i}$. 
While $K_{d}$ and $K_{i}$ as equilibrium constants may be compared if derived from multiple binding assays, dependence of $IC_{50}$ values on experimental settings discourages its comparison across different assays \cite{li2014comparative}.

\subsection{Hyper-parameter Optimization}
Although DeepAtom is an end-to-end data-driven approach for binding affinity prediction, some hyper-parameters are inevitably introduced especially in the input featurization process. 
To finalize the optimal data representation of protein-ligand complex for DeepAtom prediction, we implement systematic optimization experiments over the related hyper-parameters. 
In all of following comparison experiments, our deep learning models are trained on the \enquote{training set-1} with \num{3390} complexes, and evaluated on the corresponding validation set with \num{377} complexes; the prediction performance is measured by Pearson's R and RMSE. 

\begin{table}[t]
\caption{Validation performance with various feature/atom types.}
\label{tb:atom_types}
\renewcommand{\arraystretch}{1.1}
\setlength{\textfloatsep}{1.1cm}
% \setlength\abovecaptionskip{-0.7\baselineskip}
% The resolution of grid is set as 1.0 \AA\ and occupancy type uses the binary representation.}
\centering
\begin{tabular}{c|c|c|c|c}
\hline
Num of Features & Resolution           & Occupancy               & RMSE  & R \\ \hline
11              & \multirow{3}{*}{1.0} & \multirow{3}{*}{Binary} & 1.485 & 0.706       \\
24              &                      &                         & 1.360 & 0.737       \\
60              &                      &                         & 1.359 & 0.737       \\ \hline
\end{tabular}
\vspace{-1em}
\end{table}

\subsubsection{Feature/Atom Types}
We consider three different descriptors with \num{11}, \num{24}, and \num{60} features. 
The first descriptor characterizes both the protein and ligand atoms with the same 11 Arpeoggio atom types.
The second descriptor is described in Section~\ref{sec:atom_types}.
The third descriptor includes 40 ANOLEA atom types to describe protein atoms and 9 element types to describe ligand atoms, as well as 11 Arpeggio atom types. 
The ANOLEA atom types describe each protein atom based on its bond connectivity, chemical nature, and whether it belongs to side-chain or backbone of the amino acid \cite{melo1997novel}\cite{melo1998assessing}.
For the rest of controlled variables, we use the simple binary scheme to represent the occupancy types, and set the resolution of 3D grid box as 1.0 \AA. 
From Table~\ref{tb:atom_types}, we can observe that when both protein and ligand atoms are treated the same (the first descriptor), a lower performance is obtained; training the models on PDBbind dataset needs extracting the structures of free protein and free ligand from the complex, assuming that the conformational change upon ligand binding is negligible. 
Therefore, binding affinity prediction relies on the inter-molecular interactions between protein and ligand atoms, while the intra-molecular energies are cancelled out. 

\begin{table}
\caption{Validation performance with different resolutions.}
\label{table: resolutions}
\renewcommand{\arraystretch}{1.1}
\centering
% The feature/atom types use the optimal scheme with 24 features, and occupancy type uses the binary representation.}
\begin{tabular}{c|c|c|c|c}
\hline
Num of Features & Resolution & Occupancy & RMSE & R \\ \hline
\multirow{2}{*}{24} & 0.5 & \multirow{2}{*}{Binary} & 1.357 & 0.739 \\
 & 1.0 &  & 1.360 & 0.737 \\ \hline
\end{tabular}
% \vspace{-1em}
\end{table}

\subsubsection{Resolution}
High-resolution rasterized data can adequately capture the fine-grained features and changes in the local spatial regions. 
However, it will cause excessive memory usage and heavy computational cost. 
Thus, there exists a trade-off between prediction performance and computational efficiency. 
Based on our analysis, we pick the resolution as 1.0 \AA\ and 0.5 \AA, both of which are less than the smallest $2\times r_{vdw}$ value of 1.4 \AA\ for the 9 major heavy atoms. 
Table~\ref{table: resolutions} shows that with the increase of resolution, DeepAtom prediction performance improves. 
However, this slight improvement comes with a large increase in the computational cost, especially when the more demanding occupancy strategy such as PCMax is utilized later; therefore we select 1.0 \AA\ as the optimal resolution value.

\begin{table}
% \vspace{-1em}
\caption{Validation performance with different occupancy types.}
\label{table: occupancy types}
\renewcommand{\arraystretch}{1.1}
\centering
% The feature/atom types use the optimal scheme with 24 features, and resolution of 3D gird is set as the optimal value 1.0 \AA\.}
\begin{tabular}{c|c|c|c|c}
\hline
Num of Features     & Resolution           & Occupancy & RMSE  & Pearson\textquotesingle s R \\ \hline
\multirow{2}{*}{24} & \multirow{2}{*}{1.0} & Binary    & 1.360 & 0.737       \\
                    &                      & PCMax     & 1.348 & 0.741       \\ \hline
\end{tabular}
\vspace{-1.5em}
\end{table}

\subsubsection{Occupancy Type}
Occupancy type describes how each atom impacts its surrounding environment. 
Several different strategies have been proposed, such as binary, Gaussian\cite{ragoza2017protein} and PCMax\cite{jimenez2017deepsite}. 
The binary occupancy discretizes the atom impact over the voxel. 
For example, if the distance between an atom and a voxel center is shorter than the atom's van der Waals radius, the corresponding voxel channel will be activated as 1, otherwise deactivated as 0. 
In contrast, the Gaussian and PCMax approaches can represent an atom's impact by a continuous numerical value, which can contain richer information. The impact can also decay smoothly when the distance increases.
We compare the binary and PCMax occupancy types, on the basis of the optimal 24 feature/atom types and 1.0 \AA\ grid resolution. 
% The performance of each occupancy type is presented in the table~\ref{table: occupancy types}. 
Table~\ref{table: occupancy types} shows that DeepAtom with PCMax occupancy types achieves better performance. 
Considering the similarity between Gaussian and PCMax algorithms, we expect them yield comparable results. 

\subsubsection{Averaging at the Testing}
As an effective strategy, data augmentation is also used to improve the DeepAtom performance. 
In addition to augmenting data for training, we also run the trained model on multiple augmented versions of test data and average the results to reduce the prediction variance. 
We evaluate multiple test data versions, including 1, 12 and 24, where the value 1 means only the original test set is used without the averaging operation. 
We observe that increasing the test set versions can favorably reduce the variance of predictions and further improve the performance. 

\section{Conclusion}
In this paper, we proposed the framework DeepAtom to accurately predict the protein-ligand binding affinity. 
An efficient 3D-CNN architecture is proposed to effectively improve the model learning capacity with limited available complexes data for training.
In a purely data-driven manner without \textit{a priori} functional form assumptions, DeepAtom significantly outperforms state-of-the-art deep learning, machine learning and conventional scoring techniques.
We also proposed a new benchmark dataset to further improve the model performance. 
The promising results on independent challenging datasets demonstrated DeepAtom can be potentially adopted in computational drug development protocols such as molecular docking and virtual screening.

\section{Acknowledgments}
This work is partially supported by National Science Foundation (CNS-1842407) and National Institutes of Health (R01GM110240).
We thank Yaxia Yuan for helpful discussions and comments that improved the manuscript.

\bibliographystyle{abbrv}
\bibliography{Refs.bib}

\end{document}